\pdfoutput=1
\documentclass{JINST}
\usepackage{wrapfig}
\usepackage{subfig}
\usepackage{textgreek}

\title{MCP-based Photodetectors for Cryogenic Applications}

\author{Ranjan Dharmapalan$^a$\thanks{Corresponding
author.}~, Anil Mane$^b$, Karen Byrum$^a$,  Marcel Demarteau$^a$, Jeffrey Elam$^b$, Edward May$^a$, Robert Wagner$^a$, Dean Walters$^a$, Lei Xia$^a$,  Junqi Xie$^a$, Huyue Zhao$^a$ and J. Wang$^a$\\
\llap{$^a$}HEP Division, Argonne National Laboratory,\\
 Argonne, IL-60439, USA,\\
\llap{$^b$}Energy Systems Division, Argonne National Laboratory,\\
 Argonne, IL-60439, USA\\
  E-mail: \email{rdharmapalan@anl.gov}}

\abstract{
The Argonne MCP-based photo detector is an offshoot of the Large Area 
Pico-second Photo Detector (LAPPD) project, wherein 6 cm x 6 cm sized detectors 
are made at Argonne National Laboratory. 
We have successfully built and tested our first detectors for pico-second timing and few mm spatial resolution.
 We discuss our efforts to customize these detectors to operate in a cryogenic environment. Initial
 plans aim to operate in liquid argon. 
We are also exploring ways to mitigate wave length shifting requirements and also 
developing bare-MCP photodetectors to operate in a gaseous cryogenic environment.}


\begin{document}

\section{Introduction}

        The Large Area Pico-second Photodetector (LAPPD) [1] project is a US Department of Energy (DOE) funded initiative to develop the next generation large area photodetectors. In order to achieve this goal in a cost-effective manner LAPPD has developed low cost, commercially viable methods to fabricate 400 cm2 thin form-factor multi channel plate photomultiplier tubes (MCP-PMTs) functionalized by Atomic Layer Deposition (ALD) [2]. The MCP substrates are made using borosilicate glass by Incom Inc. and the ALD is performed at Argonne National laboratory (ANL). The resistive and secondary emission properties of the MCP-PMTs is imparted by the ALD process [3]. Additionally ALD offers the ability to engineer the material properties and optimize the MCP-PMT performance for various applications.

\subsection{Cryogenic Applications of MCP-PMT Photodetectors}

A number of current and upcoming neutrino and dark matter search experiments employ noble liquid or gases as the detector medium [4,5]. Operating at cryogenic temperatures, these experiments rely on scintillation light in the vacuum ultra violet (VUV) regime, which is wavelength shifted (WLS), from particle interactions in the medium to infer time and/or position of interaction, crucial for reconstruction and background rejection. In particular, for neutrino experiments using liquid Argon time projection chamber, an enhanced light detection system can help with a number aspects of physics analyses [6].

\begin{figure}
\includegraphics{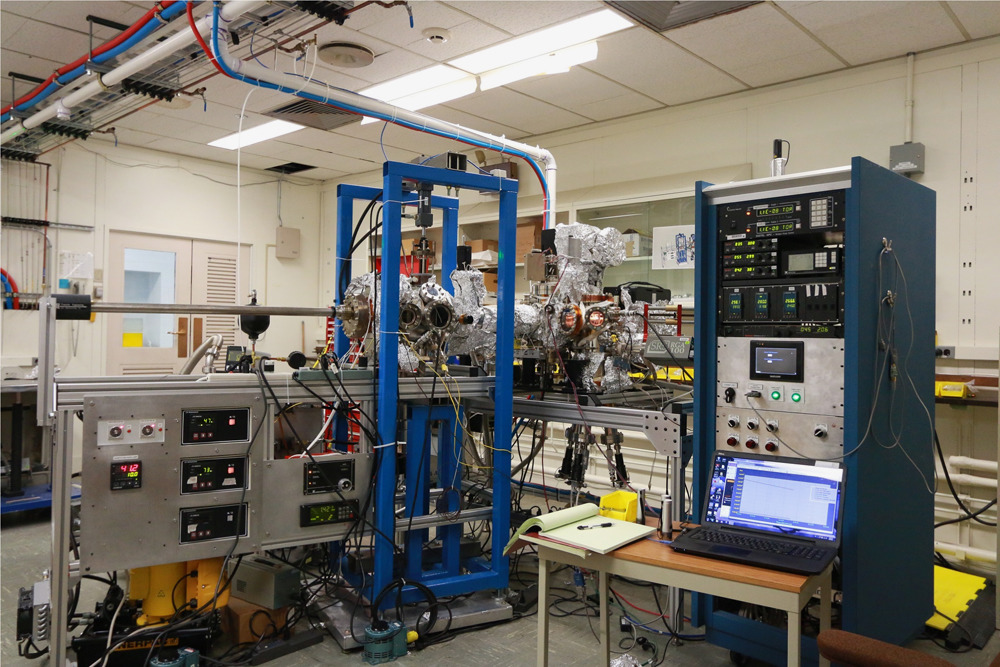}
\includegraphics{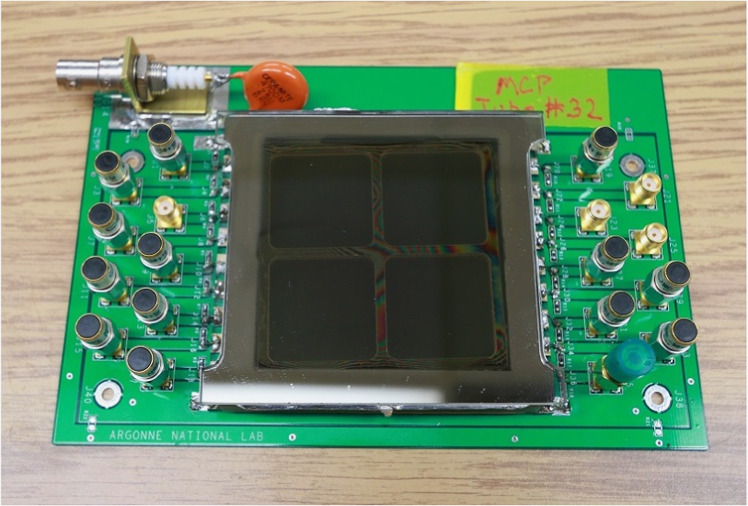}
\caption{The ANL MCP-PMT photodetector processing system.}
\end{figure}

\section{Argonne MCP-PMT Photodetector}
Argonne National Laboratory has built a photodetector processing system [see figure 1] for producing MCP-PMT based photodetectors with an active area of 6 cm x 6 cm. It is both an R\&D facility  and an intermediate step towards production and commercialization of the larger 20 cm x 20 cm photodetector [7]. A number of issues related to the manufacture of MCP-PMT photodetectors  were   addressed in the ANLprocessing system, such as, feasibility of using indium-based hermetic sealing, and outgassing of MCPs. Recently a number of long-lived prototype detectors were produced which have been characterized at ANL and also sent out to early adopters for testing.

\begin{figure}
\centering
\includegraphics{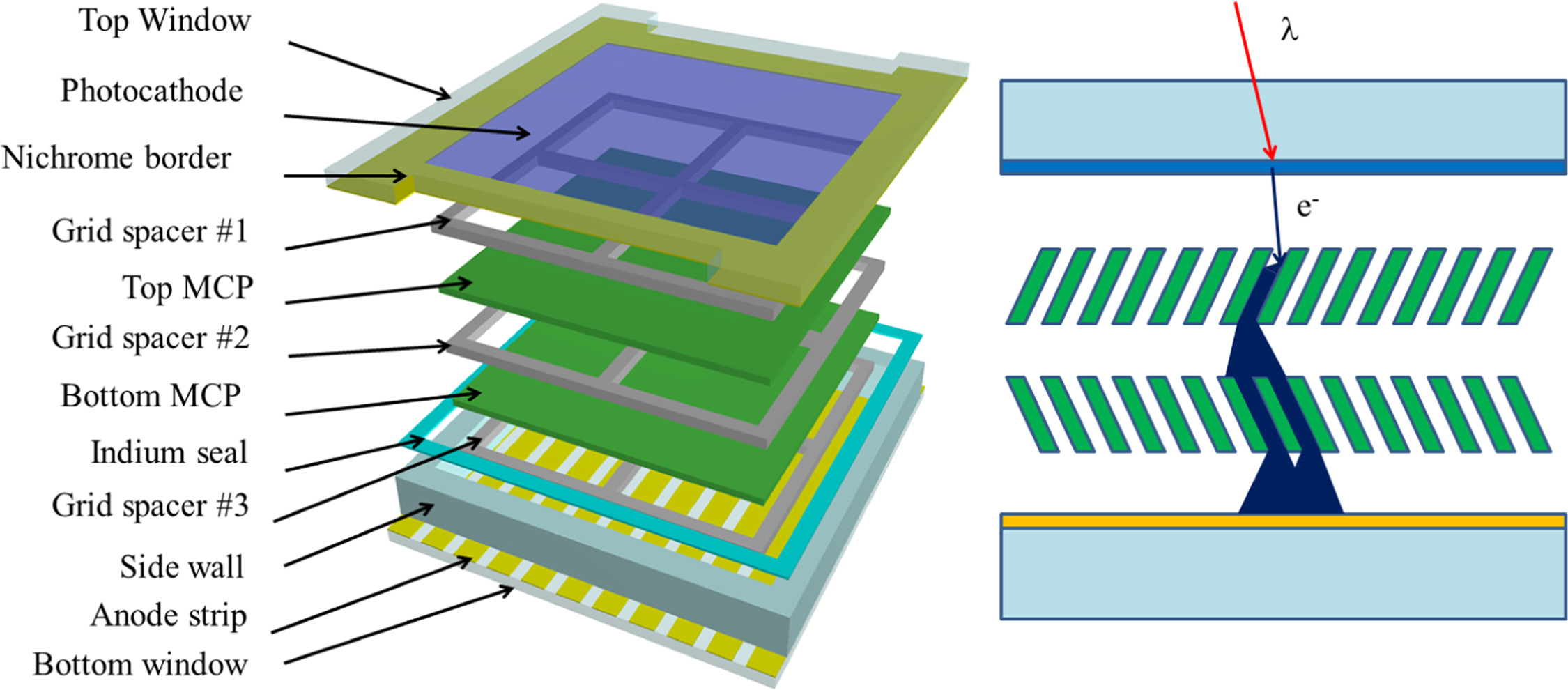}
\caption{An `exploded' view of the ANL MCP-PMT photodetector along with an illustration of its principle of opreration.}
\end{figure}

\subsection{Design of Argonne MCP-PMT Photodetector}

        The design of the ANL MCP-PMT is shown in figure 2. It consists of a pair of  ALD coated MCPs separated by glass grid spacers. The grid spacers are also coated using the ALD process to achieve the desired resistance [8]. The `stack' of MCPs and grid spacers is enclosed in an all-glass package. The glass package includes a glass bottom plate with a silk-screened anode stipline readout [9], a glass sidewall and a glass top window with a bi-alkali (K, Cs) photocathode. The sidewall is glass-frit bonded to the bottom plate while the top window is sealed in vacuum with indium [Figure 4].

\begin{figure}
\centering
\subfloat[]{\includegraphics[height=4cm,width=4cm]{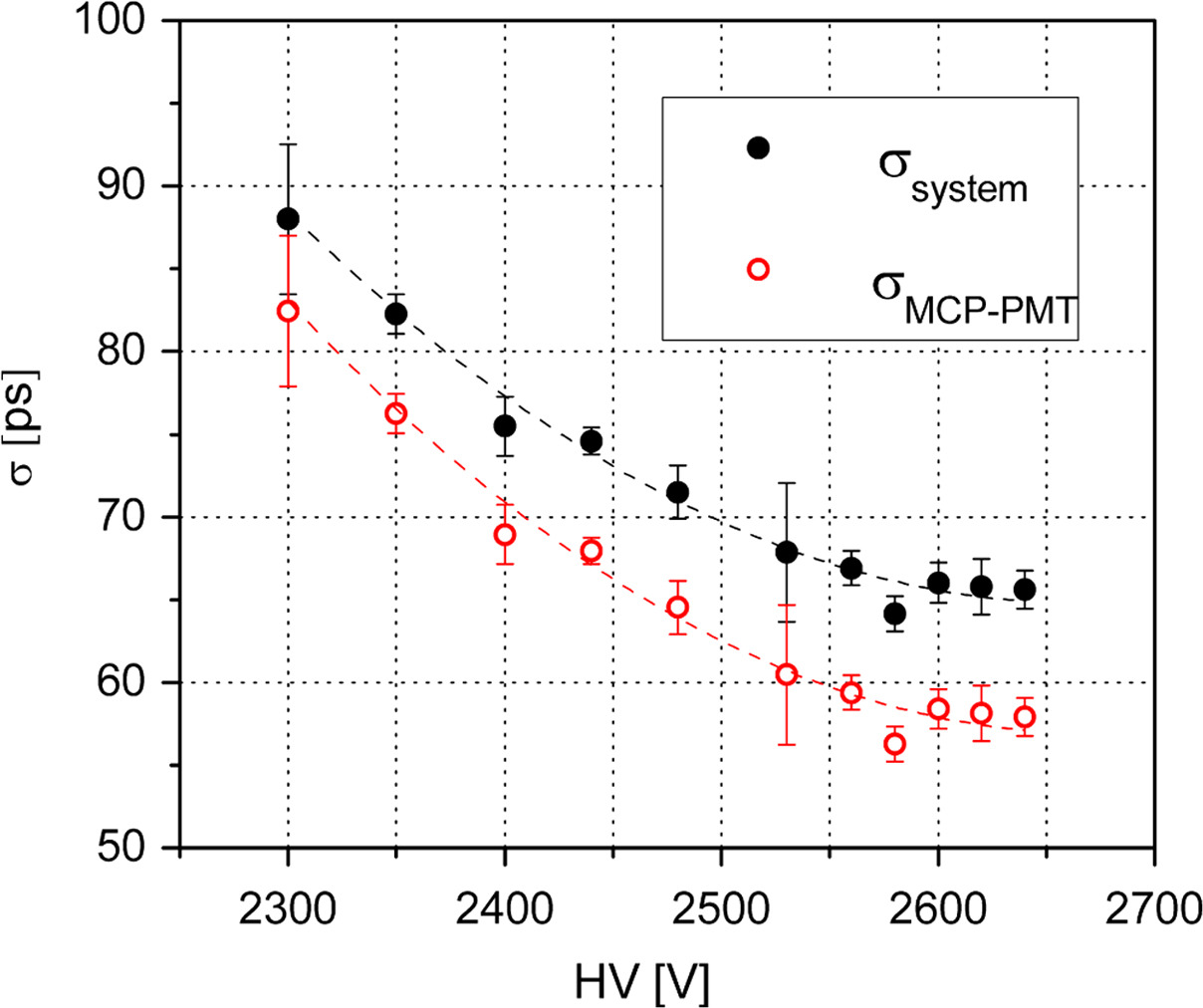}}
\subfloat[]{\includegraphics[height=4cm,width=4cm]{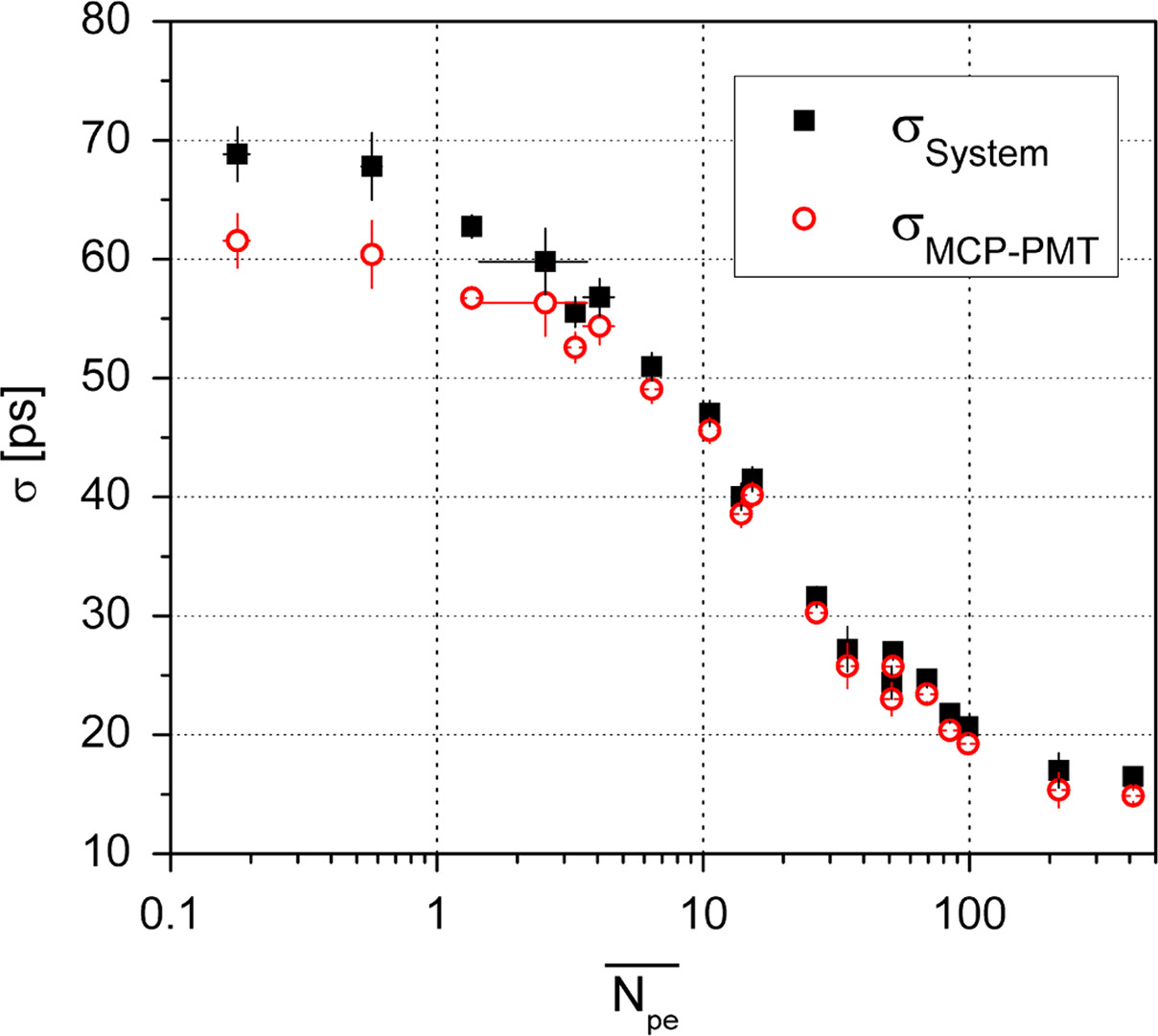}}
\hspace{3mm}%
\subfloat[]{\includegraphics[height=4cm,width=4cm]{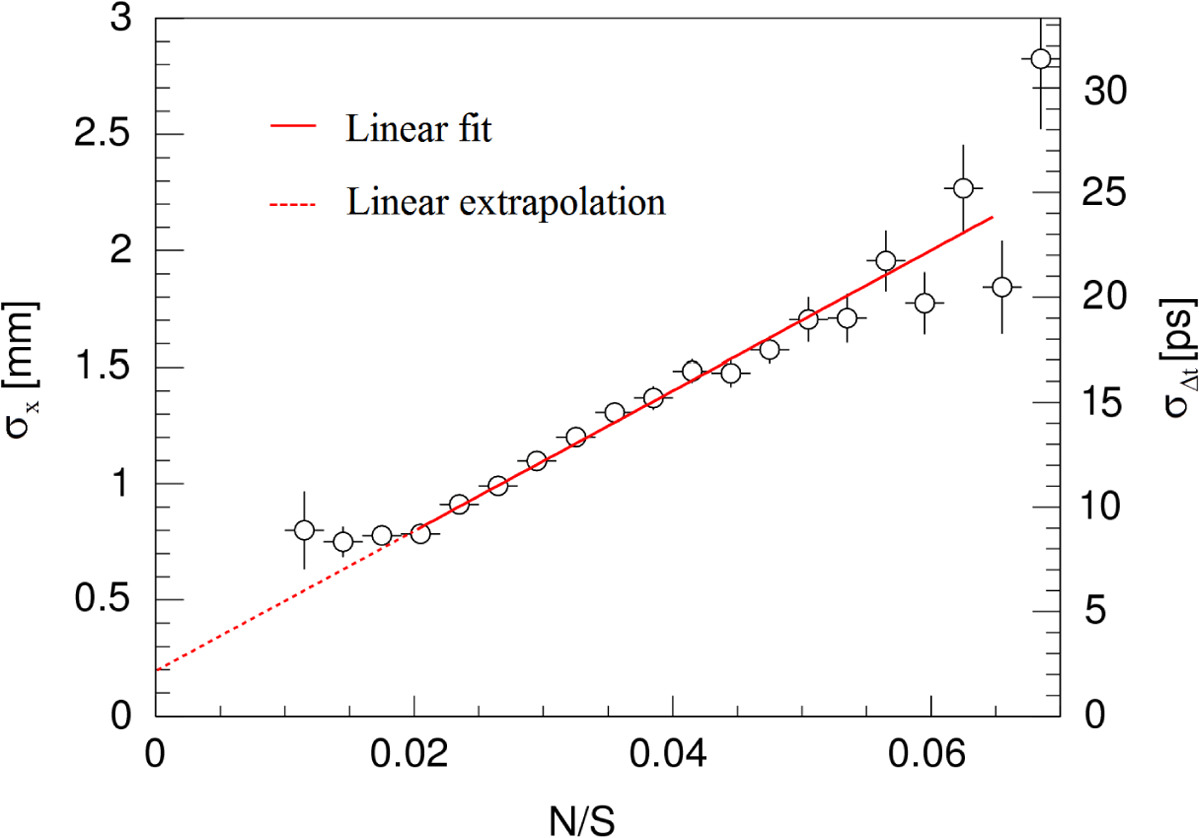}}
\caption{Highlights from ANL MCP-PMT photodetector performance tests. The transit time spread (TTS) as a function of applied
HV is show in (a). Solid circles: system time resolution. Open circles: MCP-PMT time resolution with the laser and electronics
contributions subtracted. In (b) transit time spread (TTS) as a function of the average number of photoelectrons is shown. Position
resolution (upper bound) as a function of the noise-to-signal ratio is shown in (c). The left axis is the position resolution; the right
axis is the corresponding differential time resolution. For details see Ref. [12].}
\end{figure}

\subsection{Performance of MCP-PMT Photodetectors}

        The quantum efficiency for the photocathode has been shown to reach 15\% at $\lambda \sim$ 350 nm [10].  For the MCP-PMTs made with a 20 um pore size, transit time spread resolution of 57 ps for single photoelectrons and 15 ps for multi-photoelectrons have been recorded [Figure 3(a,b)]. Position resolution of 0.75 mm was measured, including limitations due to the beam size and electronics contributions [Figure 3(c)]. These measurements were performed in a laser test facility [11] at ANL. For a detailed account of the design and testing of Argonne 6 cm x 6 cm MCP-PMT photodetector see reference [12].

\section{Cryogenic Application of Argonne MCP-PMT}

        The Argonne MCP-PMT photodetector processing system cum R\&D platform, and the fact that  the MCP substrates are ALD coated at Argonne, allows for a unique opportunity to design a photodetector  to operate in cryogenic environment. We can tune individual components  to achieve optimum performance at cryogenic temperature before manufacturing a fully sealed   cryogenic MCP-PMT photodetector.

\subsection{Testing the Sealing Technique}

\begin{wrapfigure}{l}{0.5\textwidth}
  \begin{center}
    \includegraphics[width=0.4\textwidth]{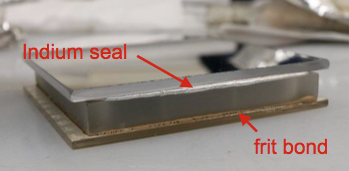}
  \end{center}
  \caption{An ANL MCP-PMT photodetector with the locations of indium seal and the glass-grit bond shown.}
\end{wrapfigure}

As a first step we tested the durability of the hermetic sealing technique developed at Argonne. The indium press seal on the top window and the glass-frit bond on the lower tile base remained intact after a number of cryogenic cycling tests in liquid nitrogen (77 K) [5].  We used both a mechanical prototype (MCP-PMT photodetector with a vacuum pump-out tube) and a fully sealed device. The device was immersed in liquid nitrogen for a maximum of 4 days. We  conducted tests with 4 devices and had one failure. The failure  due to a pre-existing hair-line crack in the glass side-wall.

\subsection{Tuning the MCP Resistance}
\begin{wrapfigure}{r}{0.5\textwidth}
  \begin{center}
    \includegraphics[width=0.4\textwidth]{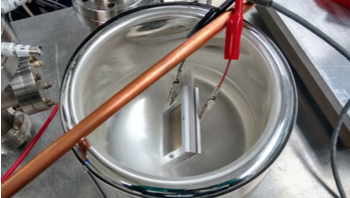}
  \end{center}
  \caption{An ANL MCP-PMT photodetector with the locations of indium seal and the glass-grit bond shown.}
\end{wrapfigure}

We investigated the resistance of the MCP  versus temperature in order to find the optimum ALD recipe for cryogenic operation [Figure 5]. Recently a few MCPs with the resistance and secondary emissive layer tuned for cryogenic operation have been produced and are being tested to measure the gain and stability. The results from these tests will be soon published elsewhere.

\subsection{Future Plans}

Going forward, we plan to make a number of MCP-PMT photodetectors engineered to operate at cryogenic temperatures. These detectors will be mounted on appropriate readout electronic boards also developed at ANL. Tetraphenyl butadiene (TPB) will be coated on to the top window  to study the response to wavelength-shifted light induced by the liquid argon scintillation signal. 
We are also testing magnesium fluoride (MgF) windows to be used as a top-window for cryogenic MCP-PMT photodetector. Along with the appropriate photocathode, this will eliminate the need for a WLS and enable direct observation of VUV scintillation light.
Another idea being pursued is to design a bare MCP-photodetector. In this design, MCP is directly exposed to cryogenic gas and the photo-conversion occurs at the MCP surface which has been ALD coated to act as a photocathode.

\acknowledgments

        Initial work on the cryogenic application of MCP-PMT photodetectors was supported by LDRD funds from ANL. We thank Joe Gregar (ANL) of the Argonne glass shop, for work on the glass-frit seal. We are deeply grateful to Matthew Wetstein (University of Chicago) and Bernhard Adams (ANL) for their advice on detector testing. Work at ANL was supported by the U.S. Department of Energy, Office of Science, Office of Basic Energy Sciences and Office of High Energy Physics under contract DE-AC02-06CH11357. Use of the Center for Nanoscale Materials was supported by the U.S. Department of Energy, Office of Science, Office of Basic Energy Sciences, under Contract no. DE-AC02-06CH11357.

\end{document}